\documentclass[11pt,a4paper]{article}
\usepackage{jcappub,epsfig,multicol,bbm,latexsym,graphicx,subfigure}
\usepackage{aas_macros}

\def\aap{Astron.\ Astrophys.\ }

\def\apjl{Astrophys.\ J.\ Lett.\ }
\def\apjs{Astrophys.\ J.\ Suppl.\ }

\begin{document}

\title{\boldmath Anisotropies of different mass compositions of cosmic rays}



\author[a,b,1]{Bing-Qiang Qiao, \note{Corresponding author.}}
\author[c,1]{Wei Liu,}
\author[c,1]{Yi-Qing Guo,}
\author[a,b,d,1]{Qiang Yuan}



\affiliation[a]{Key Laboratory of Dark Matter and Space Astronomy, Purple Mountain Observatory, Chinese Academy of Sciences, Nanjing 210008, China}
\affiliation[b]{School of Astronomy and Space Science, University of Science and Technology of China, Hefei 230026, China}
\affiliation[c]{Key Laboratory of Particle Astrophysics, Institute of High Energy Physics, Chinese Academy of Sciences, Beijing 100049, China}
\affiliation[d]{Center for High Energy Physics, Peking University, Beijing 100871, China}

\emailAdd{qiaobq@ihep.ac.cn}
\emailAdd{liuwei@ihep.ac.cn}
\emailAdd{guoyq@ihep.ac.cn}
\emailAdd{yuanq@pmo.ac.cn}

\abstract{The spectral hardenings of cosmic ray nuclei above $\sim 200$ GV
followed by softenings around 10 TV, the knee of the all-particle spectrum
around PeV energies, as well as the pattern change of the amplitude and phase of the large-scale anisotropies around 100 TeV indicate the complexities of the origin and transportation of Galactic cosmic rays. It has been shown that nearby source(s) are most likely to be the cause of such spectral features of both the spectra and the anisotropies. In this work, we study the anisotropy features of different mass composition (or mass groups) of cosmic rays in this nearby source model. We show that even if the spectral features from the nearby source component is less distinctive compared with the background component from e.g., the population of distant sources, the anisotropy features are more remarkable to be identified. Measurements of the anisotropies of each mass composition (group) of cosmic rays by the space experiments such as DAMPE and HERD and the ground-based
experiments such as LHAASO in the near future are expected to be able to
critically test this scenario.}

\maketitle
\flushbottom

\section{Introduction}
\label{sec:intro}

It is widely postulated that cosmic rays (CRs) with energies less than
$\sim$ PeV are mainly generated by the Galactic supernova remnants (SNRs).
Through the well-known diffusive shock acceleration process inside SNRs
\cite{1977ICRC...11..132A,1977DoSSR.234.1306K,1978MNRAS.182..147B,
1978ApJ...221L..29B}, CRs are accelerated to form non-thermal power-law
spectra, ${\rm d}N/{\rm d}{\cal R} \propto \cal R^{-\nu}$, with ${\cal R}$
being the rigidity. After escaping from the acceleration sites, they undergo
frequent scatterings with the random magnetic turbulence in the Galaxy,
whose behaviours are usually described by a diffusion equation.
In the conventional propagation model, the diffusion is assumed to be
homogeneous and isotropic, with a rigidity-dependence, namely $D({\cal R})
\propto {\cal R}^\delta$, with $\delta$ being constrained from $0.3$ to
$0.6$ from the Boron-to-Carbon ratio \cite{2017PhRvD..95h3007Y}.
The propagated CR spectrum should then fall as $\phi \propto \cal
R^{-\nu-\delta}$. Aside from the diffusion, CR nuclei may also suffer
from the convection, re-acceleration as well as the fragmentation by
collision with the interstellar gas. The low-energy nuclei also lose
their energies due to the ionization and Coulomb scattering. For a
comprehensive introduction to the CR propagation in the Galaxy, one can
refer to \cite{1990acr..book.....B, 2007ARNPS..57..285S}.

The convectional CR transport picture has successfully reproduced the
observed power-law spectrum, the secondary-to-primary ratio, e.g. B/C
ratio, the diffuse gamma-ray distribution and so on. However, growing
observations challenge the conventional transport model. In recent years,
the hardening of CR spectra above a few hundred GeV/nucleon received
much attention. They were initially observed by balloon-borne calorimeter
experiments ATIC-2 \cite{2007BRASP..71..494P,2009BRASP..73..564P}, CREAM
\cite{2010ApJ...714L..89A,2011ApJ...728..122Y}, and get confirmed by
precise measurements of space magnetic spectrometer experiments PAMELA
\cite{2011Sci...332...69A} and AMS-02 \cite{2015PhRvL.114q1103A,
2015PhRvL.115u1101A} and calorimeter experiment CALET
\cite{2019PhRvL.122r1102A}. The finding of the spectral hardenings
brings about various alternatives to the traditional CR paradigm (e.g.
\cite{2011ApJ...729L..13O,2012PhRvL.108h1104M,2011PhRvD..84d3002Y,
2012PhRvL.109f1101B,2012ApJ...752...68V,2012ApJ...752L..13T}).
Most recently, the DAMPE observation shows clearly that the proton
spectrum further experiences a spectral softening at $\sim14$ TeV
\cite{2019arXiv190912860A}. Hints of such spectral features were also
found previously by CREAM \cite{2017ApJ...839....5Y} and NUCLEON
measurements \cite{2018JETPL.108....5A}. These new observations indicate
that the structures of the energy spectra of CRs are more complicated
than expected (see e.g., \cite{2019arXiv190912857Y,2019arXiv191101311L}
for discussions).

In addition to the unexpected structures in CR energy spectra, the
traditional transport scenario also fails to explain the observed
anisotropies. Despite that the arrival directions of Galactic CRs are
highly isotropic due to their diffusive propagation in the Galactic
magnetic field, a weak dipole-like anisotropy is consistently observed,
with intensity differences up to $\sim 10^{-4} - 10^{-3}$. So far a
large amount of observations of anisotropies ranging from TeV to PeV
have been carried out by the ground-based experiments, for example
Super-Kamiokande \cite{2007PhRvD..75f2003G}, Tibet \cite{2005ApJ...626L..29A,
2006Sci...314..439A, 2010ApJ...711..119A, 2017ApJ...836..153A},
Milagro \cite{2008PhRvL.101v1101A, 2009ApJ...698.2121A},
IceCube/Ice-Top \cite{2010ApJ...718L.194A, 2011ApJ...740...16A,
2012ApJ...746...33A, 2013ApJ...765...55A, 2016ApJ...826..220A},
ARGO-YBJ \cite{2013PhRvD..88h2001B, 2015ApJ...809...90B} and HAWC
\cite{2014ApJ...796..108A}. Besides the large-scale anisotropies, some
mediate-scale and small-scale anisotropies have also been measured
\cite{2014ApJ...796..108A,2016ApJ...826..220A,2019ApJ...871...96A}.

One of the well-known origins of the large-scale anisotropy is the so-called
Compton-Getting effect \cite{1935PhRv...47..817C, 1968Ap&SS...2..431G},
which is induced by the motion of solar system with respect to a frame
in which CR distribution is isotropic. This effect only relies on the
power-law index of CR energy spectrum as well as the velocity of solar
system, and does not vary with energy. The diffusion of CRs also predicts
a large-scale dipole anisotropy, whose amplitude is expected to be
proportional to the diffusion coefficient and the phase is along the
density gradient of CRs. However the observations show a more complicated
energy dependence. Less than $100$ TeV, the amplitude of anisotropy
increases first with energy and then decreases, and is far below the
prediction of the standard diffusion scenario \cite{2005JPhG...31R..95H,
2006APh....25..183E,2006AdSpR..37.1909P,2012JCAP...01..011B}.
Furthermore, the phase also disagrees with the observations.
The conventional propagation model predicts the excess of CR flux toward
the direction of the Galactic center. However, the TeV measurements show
the excess approaches to the direction of the heliotail, i.e. so-called
tail-in region \cite{1998JGR...10317429N}. The puzzle is commonly
referred to as the ``anisotropy problem". The anisotropy problem may
indicate the effects due to the regulation of local magnetic field
and/or nearby sources \cite{2014Sci...343..988S,2015ApJ...809L..23S,
2015PhRvL.114b1101M,2016PhRvL.117o1103A,2013APh....50...33S,
2017PhRvD..96b3006L}.

It has been noted that there exists a common energy scale between the
structures of the energy spectra and the large-scale anisotropies,
which may indicate a common origin of them \cite{2019JCAP...10..010L}.
We proposed in a recent work that these features might be the imprints
of local sources \cite{2019JCAP...10..010L}. The spectral softenings
around 10 TeV are due to a nearby source contribution on top of the
background component. The low-energy ($\lesssim100$ TeV) anisotropies
are dominated by the local source, whihe the high-energy anisotropies
are due to the background. The transition of the low-energy and
high-energy components occur at about 100 TeV, forming a dip in the
amplitude and a flip of the phase from nearly anti-Galactic center
direction to the Galactic center direction. In \cite{2019JCAP...10..010L},
only protons and Helium nuclei are considered. In this work, we further
extend this model to heavier nuclei. We pay particular attention to
the anisotropy features of different mass composition (or mass groups),
which may be tested in the near future by e.g., LHAASO
\cite{2010ChPhC..34..249C,2019arXiv190502773B}.

\section{Model Description}

\subsection{Spatially-dependent diffusion}

We work in a spatially-dependent propagation (SDP) frame, which is
motivated by the HAWC observations of extended haloes around pulsars
\cite{2017Sci...358..911A}. The SDP model was proposed to account for
the hundred GeV spectral hardenings of CRs \cite{2012ApJ...752L..13T,
2016ApJ...819...54G,2016ChPhC..40a5101J}. It was then found to be
able to explain a series of observations of CR spectra and diffuse
$\gamma$-rays \cite{2018PhRvD..97f3008G,2018ApJ...869..176L}.
The diffusion volume in the SDP model is separated into two regions.
Close to the Galactic disk $(|z| < \xi z_h)$, where $z_h$ is the half
thickness of the whole diffusive halo, the level of turbulence is
expected to be high due to activities of supernova explosions, and
hence the diffusion coefficient is relatively small. In the outer
halo $(|z| > \xi z_h)$, particles diffuse much faster. The parameterized
diffusion coefficient we adopt is \cite{2018PhRvD..97f3008G,
2018ApJ...869..176L}
\begin{equation}
D_{xx}(r, z, \mathcal R) = D_0 F(r, z) \beta^\eta \left(\frac{\mathcal R}{\mathcal R_0} \right)^{\delta_0 F(r, z)} ~,
\end{equation}
where
\begin{equation}
 F(r,z) =
\begin{cases}
g(r,z) +\left[1-g(r,z) \right] \left(\dfrac{z}{\xi z_0} \right)^{n} , &  |z| \leq \xi z_0 \\
1 ~, & |z| > \xi z_0
\end{cases},
\end{equation}
with $g(r,z) = N_m/[1+f(r,z)]$, and $f(r,z)$ is the source density
distribution. The numerical package DRAGON \cite{2008JCAP...10..018E}
is used to solve the transport equation.
In this work, we adopt the diffusion-reacceleration model.

The injection spectrum of background sources is assumed to be a power-law
of rigidity with a high-energy exponential cutoff, $q({\cal R}) \propto
{\cal R}^{-\nu} \exp(-{\cal R}/{\cal R}_{\rm c})$. The cutoff rigidity
of each element could be either $Z$- or $A$-dependent. The spatial
distribution of sources takes the form of SNR distribution
\cite{1996A&AS..120C.437C}, $f(r,z) \propto (r/r_\odot)^{1.69}
\exp[-3.33(r -r_\odot)/r_\odot] \exp(-|z|/z_s)$,
where $r_\odot = 8.5$ kpc and $z_s = 0.2$ kpc.




\subsection{Local source}

The time-dependent propagation of CRs from the local source is obtained
using the Green's function method, assuming a spherical geometry with
infinite boundary conditions. The solution is
\begin{equation}
\phi(r, {\cal R}, t) = \dfrac{q_{\rm inj}({\cal R})}{(\sqrt{2\pi} \sigma)^3} \exp \left(-\dfrac{r^2}{2\sigma^2} \right) ~,
\end{equation}
where $q_{\rm inj}({\cal R})\delta(t)\delta(\vec{r})$ is the instantaneous
injection spectrum of a point source, $\sigma({\cal R}, t) =
\sqrt{2D({\cal R})t}$ is the effective diffusion length within time $t$.
The diffusion coefficient $D({\cal R})$ is taken the value nearby the
solar system. The injection spectrum is again parameterized
as a cutoff power-law form, $q_{\rm inj}({\cal R})=q_0{\cal R}^{-\alpha}
\exp(-{\cal R}/{\cal R}'_{\rm c})$. The normalization $q_0$ is determined
through fitting to the GCR energy spectra. The distance and age of the
local source are set to be $d = 330$ pc and $\tau = 3.2 \times 10^5$ years
\cite{2019JCAP...10..010L}, respectively. The direction of the local source is obtained through fitting to the data of the anisotropies. We find that for $l=161^{\circ}$ and $b=9^{\circ}$, both the amplitudes and phases of the large-scale anisotropies can be reproduced (see below). We further assume that the local source contributes only to primary nuclei (such as p, He, C, O, Fe), rather than secondary nuclei (such as B and Be).

\subsection{Sun's vertical location}

It should be noted that usually the solar system is assumed
to be located at the mid-plane of the Galactic disk, and the source
distribution is symmetric above and below the disk. However, it has long
been known that the Sun locates slightly above the Galactic plane (towards
the north Galactic pole). The inferred distance above the mid-plane is from
several parsecs to $\sim20$ pc \cite{2007MNRAS.378..768J,2016AstL...42..182B,
2017MNRAS.468.3289Y}. The offset may induce a net vertical flow outwards
from the Galactic plane, which generates a corresponding component of
anisotropy. The total large-scale anisotropies thus include three
components, the radial component, the vertical component, and the local
source component. The sum of these three components give the total
anisotropies which can be used to compare with the data. The vertical
location of the Sun is assumed to be 10 pc above the Galactic mid-plane
in this work.

\section{Results}

\begin{figure*}
\centering
\includegraphics[height=6.cm, angle=0]{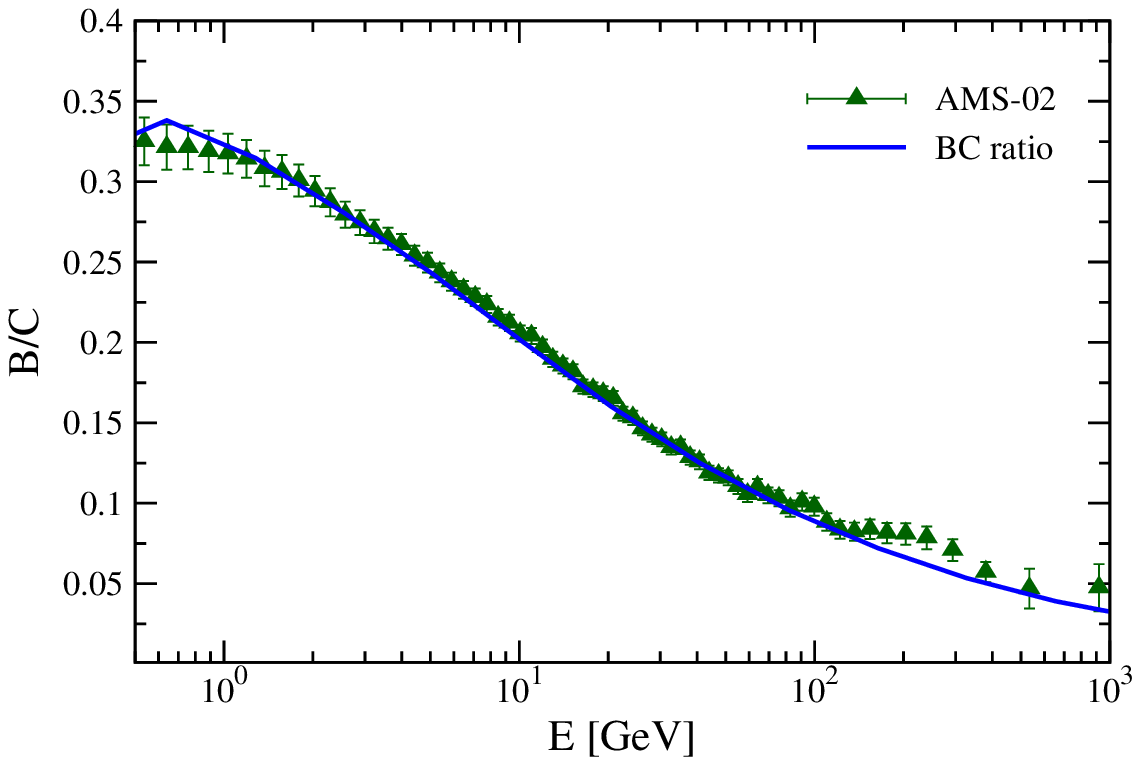}
\includegraphics[height=6.cm, angle=0]{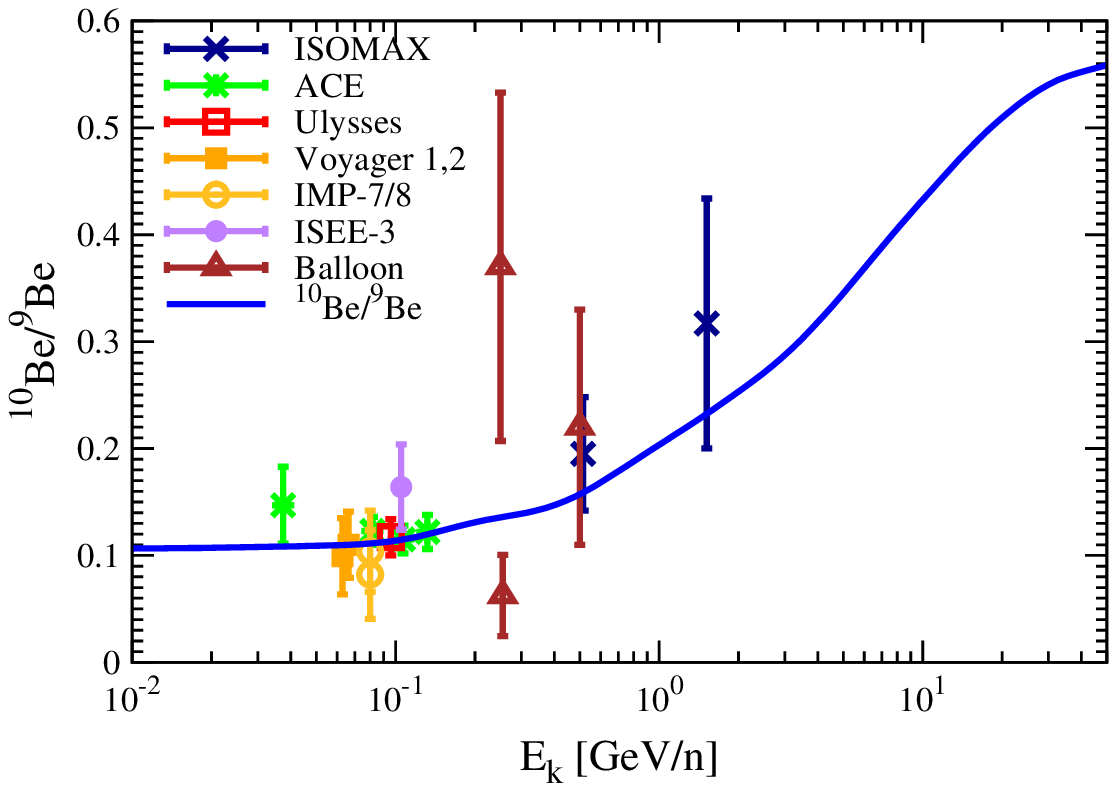}
\caption{
Model predictions of the B/C ratio (left panel) and
${}^{10}$Be/${}^{9}$Be ratio (right panel), compared with the
measurements \cite{PhysRevD.91.063508}.
}
\label{fig:ratio}
\end{figure*}

The model parameters are tuned according to the B/C and
${}^{10}$Be/${}^{9}$Be ratios, the energy spectra of various nuclei species,
the all-particle spectra, and the amplitudes and phases of the anisotropies.
The diffusion coefficient parameters are $D_0 = 8.34\times 10^{28}$
cm$^2$~s$^{-1}$, $\delta_0 = 0.65$, $N_m = 0.39$, $n = 3.5$, $\xi=0.1$,
and $\eta=0.05$. The thickness of the propagation halo is $z_h=10$ kpc,
and the Alfv\'enic velocity is $v_A=6$ km~s$^{-1}$. Note that a larger
value of $z_h$ is adopted in this work compared with
Ref.~\cite{2018PhRvD..97f3008G}. This is to suppress the vertical
component of the anisotropies which seems to be lack in the observations
at high energies ($>100$ TeV). It is interesting to note that previous
studies under the simple one-zone propagation framework also give a
relatively large value of the halo height \cite{2011ApJ...726...81A,
2013MNRAS.436.2127O}. The comparison of the B/C and ${}^{10}$Be/${}^{9}$Be
ratios between the model predictions and the data is given in Fig.
\ref{fig:ratio}.

\begin{figure*}
\centering
\includegraphics[height=14.cm, angle=0]{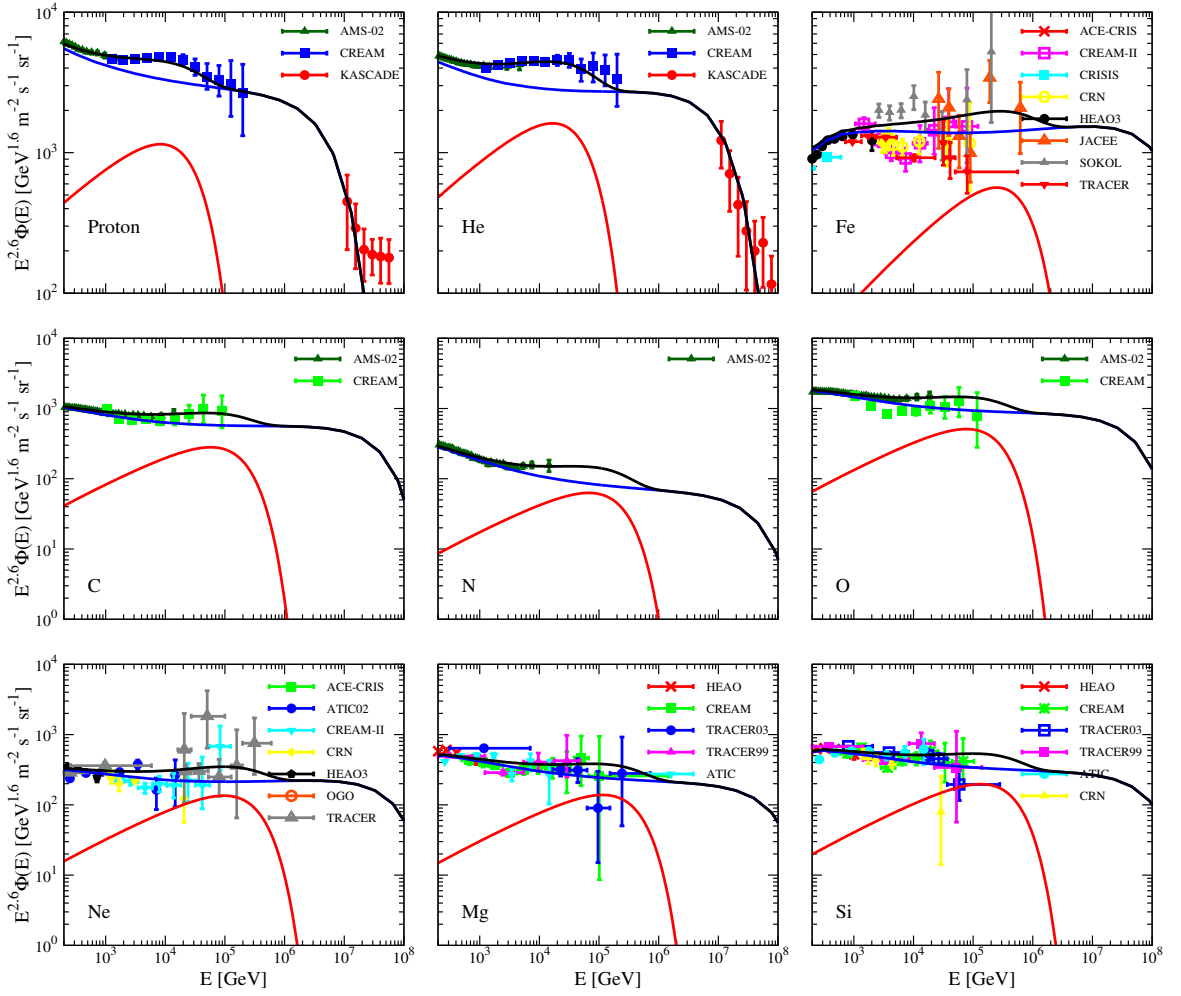}
\caption{
The computed energy spectra of protons, He, C, N, O, Ne, Mg, Si, and Fe
nuclei. Here the flux is multiplied by $E^{2.6}$. The red and blue lines
are the fluxes from the local and background sources, and the black line
is their sum.
For protons, He, C, N and O, the data points are taken from AMS-02 \cite{2015PhRvL.114q1103A, 2017PhRvL.119y1101A, 2018PhRvL.120b1101A, 2018PhRvL.121e1103A}, CREAM \cite{2009ApJ...707..593A, 2017ApJ...839....5Y}, and KASCADE \cite{2013APh....47...54A}, respectively. For Ne, Mg, Si, and Fe, the data points are from HEAO \cite{1990A&A...233...96E}, TRACER \cite{2004ApJ...607..333G, 2008ApJ...678..262A,2011ApJ...742...14O}, ATIC \cite{2009BRASP..73..564P}, JACEE \cite{2006JPhCS..47...41H,1998ApJ...502..278A}, SOKOL \cite{1993ICRC....2...17I}, CRISIS \cite{1981ApJ...246.1014Y}, ACE-CRIS \cite{2013ApJ...770..117L}, CRN \cite{1991ApJ...374..356M}.
}
\label{fig:spec}
\end{figure*}

Fig.~\ref{fig:spec} shows the propagated spectra of primary CR components,
including protons, He, C, N, O, Ne, Mg, Si, and Fe nuclei. In each panel
the blue and red lines are the contributions from the background and the
local source respectively, and the black solid line is the sum of them.
The corresponding injection parameters are given in Table \ref{tab:para_inj}.
The spectral indices of the local source component are assumed to be
slightly harder than that of the background component, which helps fit
the data better. This is reasonable due to the diversity of CR sources,
as can be inferred from the $\gamma$-ray observations of SNRs
\cite{2011APh....35...33Y}. We can see that the addition of the local
source component can simultaneously account for the spectral hardening
features at $\sim200$ GV, and the softening features at $\sim10$ TV.
The SDP model can also give a concave shape of the propagated CR spectra,
which was previously proposed to account for the spectral hardenings
\cite{2018PhRvD..97f3008G,2018ApJ...869..176L}. However, in this work
the SDP-induced spectral hardenings do not specifically correspond to
the measured hardenings. Nevertheless, the SDP model is still necessary
in suppressing the anisotropies as will be shown below.

\begin{figure*}
\centering
\includegraphics[height=9.cm, angle=0]{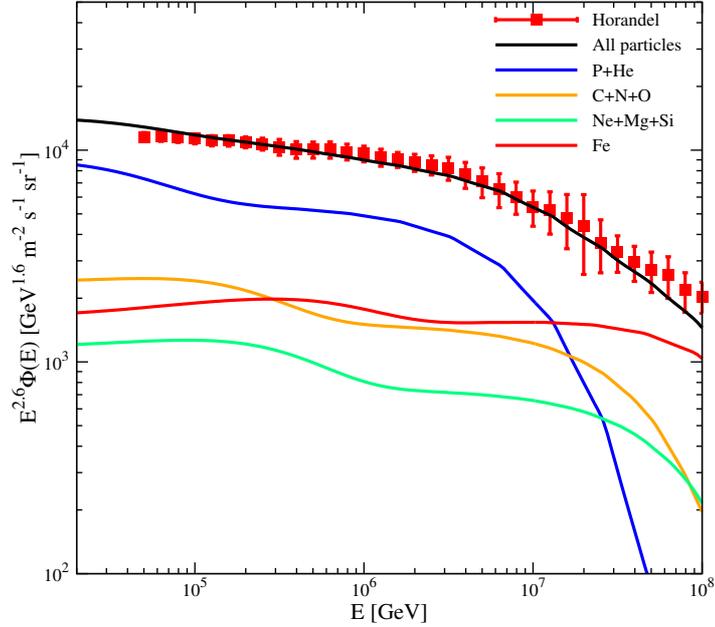}
\caption{Model prediction of the all-particle spectra, compared with
the weighted data \cite{2003APh....19..193H}.
}
\label{fig:all_spec}
\end{figure*}

Through adding different compositions together, we get the all-particle
spectrum as shown in Fig.~\ref{fig:all_spec}, compared with the weighted
data \cite{2003APh....19..193H}. The knee structure of the all-particle
spectrum can be properly reproduced by the background component assuming
a $Z$-dependent cutoff with ${\cal R}_c \sim 7$ PV. In this case we find
that the knee of the all-particle spectrum is mainly due to the suppression
of the light components (protons and He nuclei). This is because we try
to fit the KASCADE spectra of protons and He \cite{2013APh....47...54A}.
If alternatively the light component spectra from the Tibet experiments
\cite{2015PhRvD..92i2005B} are used, a smaller cutoff rigidity would be
obtained \cite{2018ChPhC..42g5103G}.

\begin{table*}
\begin{center}
\begin{tabular}{|c|ccc|ccc|}
   \hline
    & \multicolumn{3}{c|}{Background} & \multicolumn{3}{c|}{Local source} \\
   \hline
Element & Normalization$^\dagger$ & ~~~$\nu$~~~  & ~~~$\mathcal R_{c}$~~~ & ~~~$q_0$~~~~~ & ~~~~~$\alpha$~~~ & ~~~${\cal R}'_c$~~~ \\
   \hline
  & $[({\rm m}^2\cdot {\rm sr}\cdot {\rm s}\cdot {\rm GeV})^{-1}]$ & & [PV] & [GeV$^{-1}$] & &  [TV] \\
     \hline
   p   & $8.04\times 10^{-5}$    & 2.45   &  7  & $2.72\times 10^{52}$  & 2.10 & 28 \\
   He & $5.47\times 10^{-5}$   & 2.39     &  7  & $2.53\times 10^{52}$  & 2.10  &  28  \\
   C   & $1.12\times 10^{-5}$   & 2.40    &  7  & $4.80\times 10^{50}$    & 2.05 &  28  \\
   N   & $1.45\times 10^{-6}$   & 2.45    &  7  & $8.40\times 10^{49}$  & 2.05 &   28  \\
   O   & $2.30\times 10^{-5}$   & 2.43    &  7  & $5.50\times 10^{50}$ & 2.05  &   28  \\
   Ne & $4.46\times 10^{-6}$   & 2.38   &  7  & $1.02\times 10^{50}$ & 2.05  &   28  \\
   Mg & $9.06\times 10^{-6}$   & 2.44     &  7  & $7.80\times 10^{49}$ & 2.05  &   28  \\
   Si & $1.03\times 10^{-5}$     & 2.44   &  7  & $8.70\times 10^{49}$ & 2.05  &   28  \\
   Fe & $2.53\times 10^{-5}$    & 2.38    &  7  & $1.01\times 10^{50}$ & 2.05   &  28  \\
\hline
\end{tabular}\\
$^\dagger${The normalization is set at total energy $E = 1$ TeV.}
\end{center}
\caption{Injection parameters of the background and local source.}
\label{tab:para_inj}
\end{table*}

\begin{figure*}
\centering
\includegraphics[height=10.cm, angle=0]{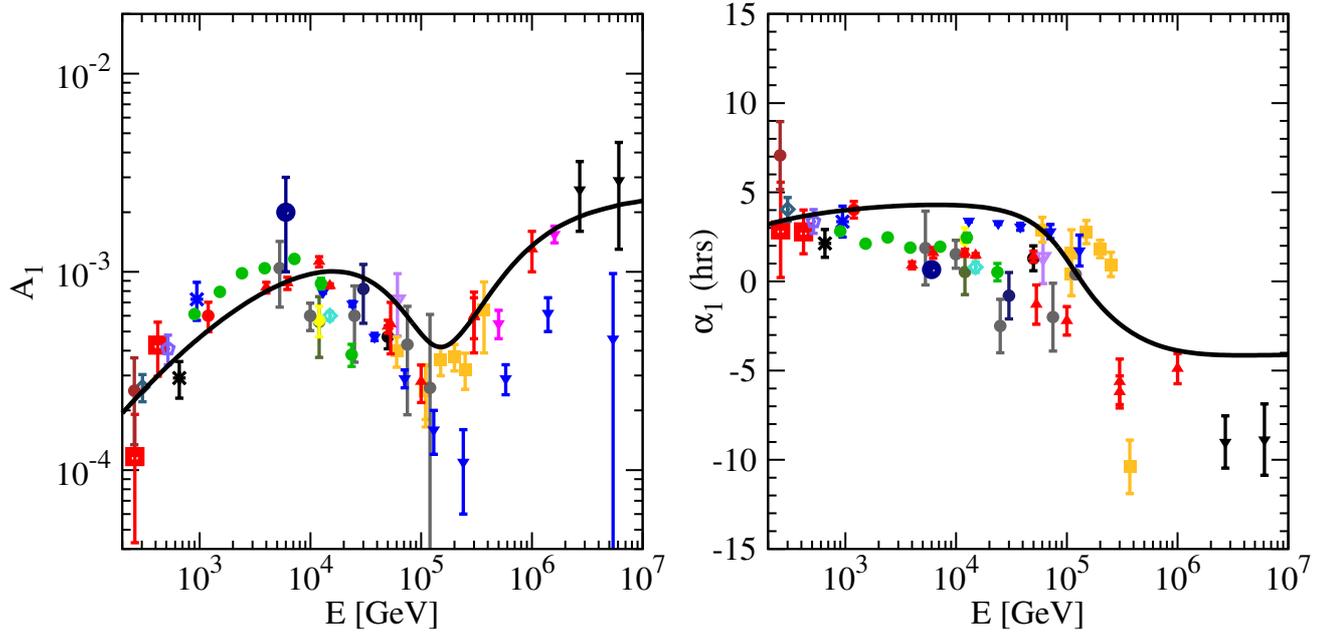}
\caption{The energy dependence of the amplitudes (left) and phases (right)
of the dipole anisotropies when adding all of the major elements together.
The data points are taken from underground muon detectors:
Norikura (1973; \cite{1973ICRC....2.1058S}),
Ottawa (1983; \cite{1981ICRC...10..246B}),
London (1983; \cite{1983ICRC....3..383T}),
Bolivia (1985; \cite{1985P&SS...33.1069S}),
Budapest (1985; \cite{1985P&SS...33.1069S}),
Hobart (1985; \cite{1985P&SS...33.1069S}),
London (1985; \cite{1985P&SS...33.1069S}),
Misato (1985; \cite{1985P&SS...33.1069S}),
Socorro (1985; \cite{1985P&SS...33.1069S}),
Yakutsk (1985; \cite{1985P&SS...33.1069S}),
Banksan (1987; \cite{1987ICRC....2...22A}),
Hong Kong (1987; \cite{1987ICRC....2...18L}),
Sakashita (1990; \cite{1990ICRC....6..361U}),
Utah (1991; \cite{1991ApJ...376..322C}),
Liapootah (1995; \cite{1995ICRC....4..639M}),
Matsushiro (1995; \cite{1995ICRC....4..648M}),
Poatina (1995; \cite{1995ICRC....4..635F}),
Kamiokande (1997; \cite{1997PhRvD..56...23M}),
Marco (2003; \cite{2003PhRvD..67d2002A}),
SuperKamiokande (2007; \cite{2007PhRvD..75f2003G});
and air shower array experiments:
PeakMusala (1975; \cite{1975ICRC....2..586G}),
Baksan (1981; \cite{1981ICRC....2..146A}),
Norikura (1989; \cite{1989NCimC..12..695N}),
EAS-TOP (1995, 1996, 2009; \cite{1995ICRC....2..800A,
1996ApJ...470..501A, 2009ApJ...692L.130A}),
Baksan (2009; \cite{2009NuPhS.196..179A}),
Milagro (2009; \cite{2009ApJ...698.2121A}),
IceCube (2010, 2012; \cite{2010ApJ...718L.194A, 2012ApJ...746...33A}),
Ice-Top (2013; \cite{2013ApJ...765...55A}),
ARGO-YBJ (2015; \cite{2015ApJ...809...90B}),
Tibet (2005, 2015, 2017; \cite{2005ApJ...626L..29A,
2015ICRC...34..355A, 2017ApJ...836..153A}).
}
\label{fig:delta_all}
\end{figure*}

As suggested in Ref.~\cite{2019JCAP...10..010L}, the softening features
in the energy spectra and the energy-dependent anisotropies might have
a common origin. In Ref.~\cite{2019JCAP...10..010L}, only the light
components of protons and Helium nuclei were considered. Here we add all
the major compositions as shown in Fig.~\ref{fig:spec} together.
The corresponding amplitudes and phases of the dipole anisotropies are
given in Fig.~\ref{fig:delta_all}. The energy dependences of both the
amplitudes and phases can be well reproduced in this model. Compared with
Ref.~\cite{2019JCAP...10..010L}, the dip of the amplitudes becomes wider,
which matches better with the data. The transition of the phase also
becomes smoother, and can be tested by improved measurements in future.
Note that the direction of the local source is different
from that assumed in Ref.~\cite{2019JCAP...10..010L}, due to the inclusion
of the vertical anisotropoies in this work.

\begin{figure*}
\centering
\includegraphics[height=10.cm,angle=0]{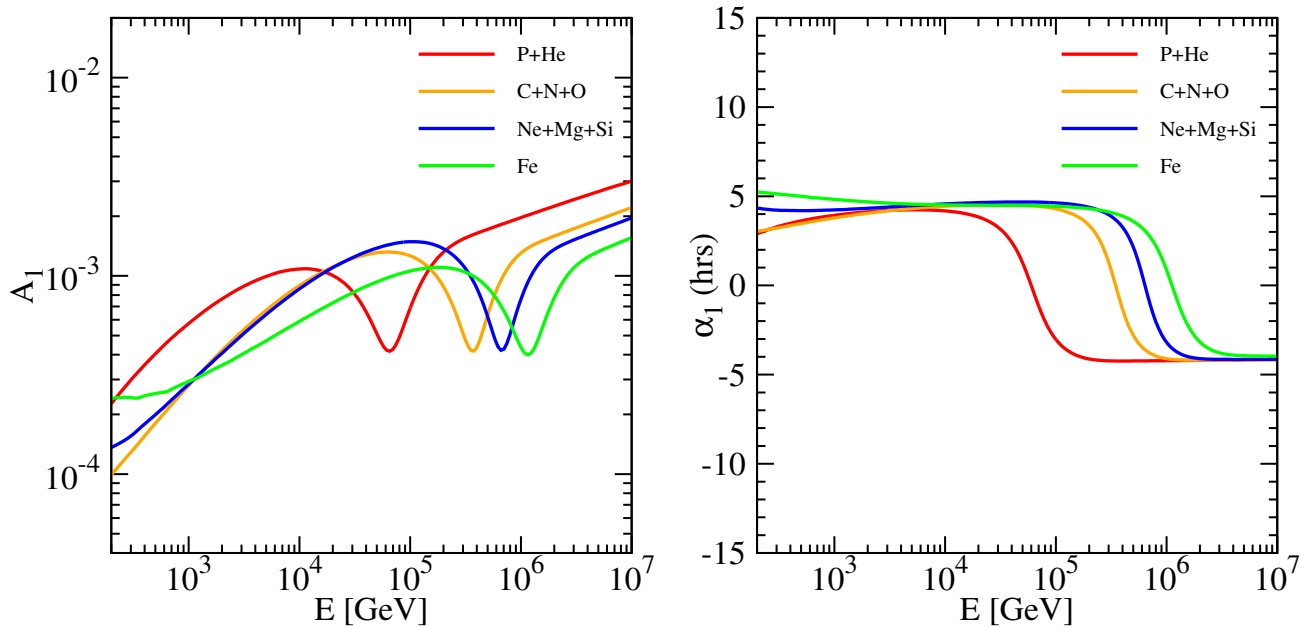}
\caption{The energy dependence of amplitudes and phases of the dipole
anisotropies for four mass groups, p+He, C+N+O, Ne+Mg+Si, and Fe, respectively.
}
\label{fig:delta_group}
\end{figure*}

We further calculate the anisotropies of different compositions.
Considering the limited particle identification capability of the
ground-based experiments, the primary components are divided into four
mass groups, i.e. p+He, C+N+O, Ne+Mg+Si, and Fe, respectively.
The resulting anisotropies are shown in Fig.~\ref{fig:delta_group}.
Dip structures of the amplitudes and phase flippings are visible for
each mass group. We expect that the observations of anisotropies of
different mass groups by LHAASO would be promising in revealing these
structures, and give a critical test of this model.

\begin{figure*}
\centering
\includegraphics[height=10.cm, angle=0]{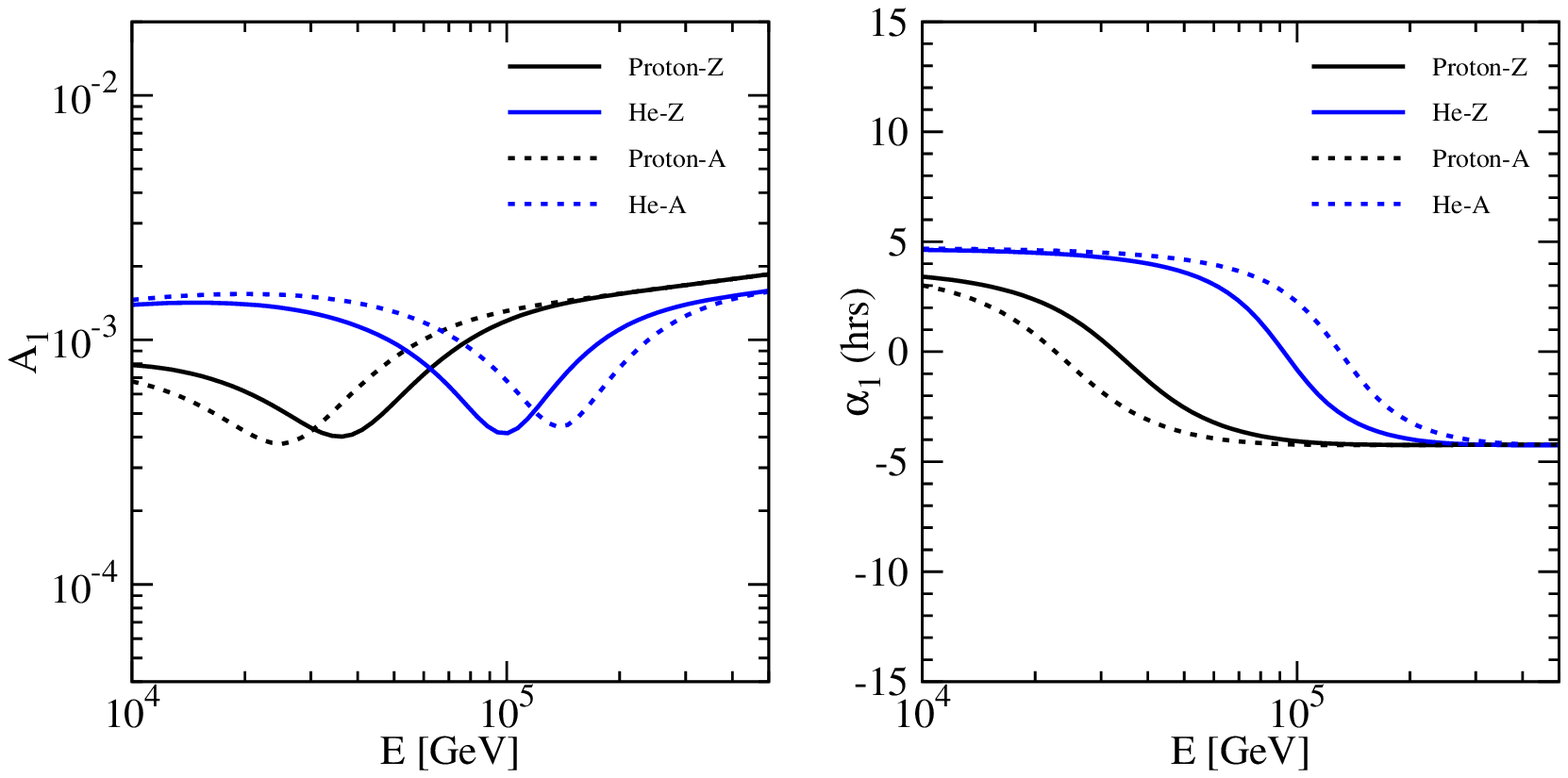}
\caption{Comparison of the amplitudes (left) and phases (right) of the
anisotropies of protons and Helium between $Z$- and $A$-dependent cutoff
of the local source spectra.
}
\label{fig:dif_ZA}
\end{figure*}

The above discussion is based on the assumption of a $Z$-dependent cutoff
energy of the local source spectra. We further investigate the effect
due to an $A$-dependent cutoff. The comparision of the anisotropy amplitudes
and phases of protons and Helium nuclei for $Z$- and $A$-dependent cutoff
are shown in Fig.~\ref{fig:dif_ZA}. For both cases, the model parameters
are tuned to fit the energy spectra of different compositions and the
total anisotropies. It is clearly shown that the energies of the dip of
protons and Helium nuclei can effectly distinguish these two assumptions.
A clear identification of protons and Helium individually is a little
bit challenging for ground-based experiments \cite{2019arXiv190409130Y}.
The measurements of anisotropies of protons and p+He are possible for
the LHAASO experiment \cite{2019arXiv190409130Y}, which can also be very
important in probing the $Z$- and $A$-dependent cutoff assumptions of
the model.

\section{Discussion}

Measurements of CRs enter a precise era thanks to fast development of
space and groundbased experiments in recent years. Based on the new
features of the CR spectra, including the spectral hardenings at $\sim200$
GV and softenings at $\sim10$ TV, together with the inhomogeneous
diffusion inferred by the HAWC observations of pulsars and the long
time puzzle of the energy-dependent evolution of the dipole anisotropy
features, an SDP frame with contributions from a local CR source was
established and could explain most of these new observational facts
\cite{2019JCAP...10..010L}.

In this work, we extend this model to study the anisotropies of heavier
nuclei. We find that the dip structure (phase flipping) of the total
amplitudes (phases) of the anisotropies becomes smoother after adding
heavier nuclei. This is because the dip features and phase changes for
different species depend on energy, and the measurements of all species
of CRs give an average effect of them. The anisotropies of different
mass groups are also investigated. Similar dip features and phase changes
are predicted for all of these compositions, with different characteristic
energies. We further explore the differences of the large-scale anisotropy
features between $Z$- and $A$-dependent assumptions of cutoff of the
local source spectra. It is expected that future precise measurements
of the anisotropies of different compositions or mass groups by e.g.,
DAMPE \cite{2017APh....95....6C}, HERD \cite{2014SPIE.9144E..0XZ},
and LHAASO \cite{2019arXiv190502773B}.

Finally we comment that the spectral features of the
electrons and positrons, particularly the remarkable positron excess
(e.g., \cite{2009Natur.458..607A,2013PhRvL.110n1102A}), can also be
properly reproduced under the same framework of the source and propagation
models as discussed in this work \cite{2019arXiv190410663T}. It is thus
very encouraging to establish a unified scenario of GCR origin and
propagation based on new precise observations of CRs and $\gamma$-rays.

\acknowledgments
This work is supported by the National Key Research and Development Program
of China (No. 2018YFA0404203), the National Natural Science Foundation of
China (Nos. 11875264, 11635011, 11761141001, 11663006, 11722328, 11851305).

\bibliographystyle{unsrt_update}
\bibliography{ref}

\end{document}